\documentstyle[osa,12pt]{revtex}

\begin{document}
\title{c-axis Tunneling in Nb/Au/YBaCuO Structures}
\author{P.V.Komissinski$^{1}$, G.A.Ovsyannikov$^{1}$, N.A.Tulina$^{2}$, V.V.Ryazanov$%
^{2}$}
\address{Institute of Radio Engineering and Electronics RAS, Moscow, Russia$^1$\\
Institute of Solid State Physics, Chernogolovka, Russia$^{2}$}
\date{25 January 1999}
\maketitle

\begin{abstract}
Abstract

We present the experimental results for Nb/Au/YBa$_{2}$Cu$_{3}$O$_{x}$
structures, in which the current flows along (001) direction of YBa$_{2}$Cu$%
_{3}$O$_{x}$ film. The theoretical evaluations show, that at the
experimental values of the Au/YBa$_{2}$Cu$_{3}$O$_{x}$ interface
transparency, determined from the interface resistance ($\bar{D}\sim 10^{-6}$%
), the critical current of the structure is of the fluctuation order of
magnitude due to the sharp decrease of the amplitude potential of the
superconducting carriers on this interface. Obtained I-V-curves could be
interpreted in terms of contact between d-type pairing superconductor or
gapless isotropic superconductor with normal metal. No critical current was
observed for investigated structures with characteristic interface
resistance $R_{N}S\sim 10^{-6}%
\mathop{\rm %
\Omega}%
\cdot 
\mathop{\rm cm}%
^{2}$, 2 orders of magnitude lower, than for known experimental data.

\end{abstract}

\section{Introduction}

The properties of high-T$_{c}$ materials are discussed on the basis of the
model of d-type symmetry of superconducting gap order parameter. This model
explains, in particular, the dependence of the critical current from the
magnetic field in bimetallic dc SQUIDs with junction YBa$_{2}$Cu$_{3}$O$_{x}$
(YBCO) and Pb [1] and the spontaneous excitation of the flux quantum in
tricrystals [2]. But some experimental results are contradictory. On the one
hand there is no critical current in the junctions YBCO/s-superconductor
along c-axis direction [3-5], which is in good agreement with the theory of
tunneling between d-superconductors and s-superconductor; on the other hand,
there are several experiments [6-8], where the evident critical current was
obtained, which amplitude value changes from magnetic and microwave fields
in accordance with the theory of s-superconductors. To explain the
experiments [6-8] it was supposed [9], that there is a mixture of d and
s-type superconducting carriers in high-T$_{c}$ materials due to their
orthorombic structure and the diffusive scattering near the interface or the
twins in high-T$_{c}$ films causes the increase of the contribution of
s-superconductor [9,10]. Note, that the evaluation barrier transparencies
(averaged over the quasiparticle momentum directions) for Pb/(Au, Ag)/YBCO
presented in [6-8] gives $\bar{D}\sim 10^{-7}\div 10^{-9}$ with the junction
areas $S=0.1\div 1mm^{2}$.

In this report we present the results of experimental investigations of
current transport in the small areas junctions ($S=8$x$8\mu m^{2}$)
s-superconductor/normal metal/high temperature superconductor (Nb/Au/YBCO)
with the averaged Au/YBCO c-axis interface transparency $\bar{D}\sim 10^{-6}$%
. The experimental results are analyzed from two points of view: classical
BCS theory and modern theoretical investigations, where d-type symmetry of
the order parameter in YBCO film is supposed.

\section{Fabrication Procedure and Experimental Samples}

The steps of the fabrication procedure are presented on the figure 1. YBCO
thin films were made with two methods: the laser ablation technique and DC
sputtering at high oxygen pressure. We used two kinds of substrates: (110)
oriented NdGaO$_{3}$ or r-cut sapphire with CeO$_{2}$ buffer layer. So, we
got c-axis oriented (001) epitaxial YBCO films with 100-150nm thickness. The
following superconducting properties were obtained: the superconducting
transition temperature T$_{cd}$=84-88K and superconducting temperature width 
$\Delta T_{cd}$=0.5-1K; resistance ratio at 300K and 100K R$_{300\text{K}}$/R%
$_{100\text{K}}\approx $2.8. The density of particles on the YBCO film
surface was $\sim 10^{6}$cm$^{-2}$. The high quality of obtained YBCO films
is confirmed by low value of the width of X-ray (005) $\Theta /2\Theta $
scan FWHM(005)$\approx 0.2^{\circ }$ for 150nm thick YBCO film.

We used the following fabrication procedure of Nb/Au/YBCO junctions: a)
Nb/Au/YBCO trilayer structure deposition; b) definition of the area of the
junction with photolithography, ion and plasma chemical techniques; c)
deposition of insulating CeO$_{2}$ layer to prevent the electrical contacts
in a-b plane of YBCO; d) Contact pads deposition and structuring.

The thin layer of the normal metal (N-metal) (Au, Ag, Pt) 10-20nm thickness
was deposited in-situ by RF sputtering after YBCO film deposition (fig.1a).
The next step is the DC-magnetron deposition of Nb counter electrode
100-150nm thickness. The absence of chemical reaction between Au and Nb
causes the using of Nb. Note, the possible chemical interaction between Au
and Pb could be occurred in the experiments [4-7], where Pb is used.

On the next step the area of the junction was defined using the
photolithography, ion and plasma-chemical etching (fig.1b). To prevent the
electrical contact in a-b plane of YBCO film, the area of the junction with
the central window 8x8$\mu m^{2}$ was insulated by CeO$_{2}$ layer using
self-aligned technique (fig.1c). At the last step the contact pads from Au
is formed(fig. 1d). We used the 2 points contact pads to the top (Nb) layer
in order to realize 4 points measurement technique at the temperature below
the T$_{c}$ of YBCO film. The junctions were positioning on the parts of
YBCO film surface, which had no particles. For measurement of and critical
current density two orthogonal 4 $\mu m$ width YBCO bridges were made on the
same substrate. More than 30 samples Nb/N-metal/YBCO with Au, Ag and Pt as a
N-metal were fabricated. Here we discuss the results of investigations of 9
samples Nb/Au/YBCO with the small tolerance of the normal resistance $R_{N}S$
(less than 4 times, see table 1) made with the same fabrication parameters.

\section{Experimental results}

The dependencies of the junction resistance from temperature $R(T)$ at bias
current 1-5$\mu A$ and I-V curves at temperature range 4.2-300K were
measured. The fig.2 presents the $R(T)$ dependencies for one of the
junctions and for YBCO bridge placed on the same substrate. The dependence $%
R(T)$ for one of the junction is in the good correspondence with one for
YBCO microbridge, which is situated on the same substrate. At $T>T_{cd}$ $%
R(T)$ has a metallic dependence, i.e. the decrease of $R(T)$. This behavior
is usual for a-b plane current flow, so YBCO\ film gives the main
contribution to the resistance at $T>T_{cd}$. The inset to the fig.2 shows
the $R(T)$ dependence of the junction at $T<T_{cd}$, the increase of the
resistance with the decrease of temperature. The $R(T)$ value is dependent
from the measurement current.

The dependencies of $R_{d}$ from the bias voltage are presented at fig.3 at
different temperatures. The value of $R_{d}(V=0)$ increase with the decrease
of $T$, which is reflected on $R(T)$ dependence. The strong nonlinear
behavior of I-V-curve at 72K%
\mbox{$<$}%
T%
\mbox{$<$}%
84K is due to a break down of superconductivity in YBCO\ film by transport
current. Note, that there is a difference between the $R_{N}=R_{d}(0)$ of
the junction near $T_{cd}$ and the asymptotic $R_{N}=R_{d}(V), V>20mV$ at $%
T<<T_{cd}$.

Parameters of the investigated junctions, fabricated at equal conditions,
are presented in table 1. The interface resistance $R_{N}S$ allows to
evaluate the averaged interface transparency, which will be used in future
[11]:

\[
\bar D=\frac{2\pi ^2\hbar ^3}{e^2p_F^2}\frac 1{R_NS}=\frac{2\rho
^{YBCO}l^{YBCO}}{3R_NS}, 
\]
where $p_{F}$ - the minimum value of Fermi-momentum from YBCO and Au. The
values of transparency for the interface Au/YBCO are presented in the table
1 for fabricated samples. We use the values $\rho ^{YBCO}l^{YBCO}=3.2\cdot
10^{-11}\Omega \cdot cm^{2},R_{N}=R_{d}(V)$ at $V=20 mV$.

We also investigated the bilayer structures Au/YBCO, Nb/YBCO and Au/Nb,
prepared with the same technology as trilayer ones. The resistance of these
interfaces are as follows: $R_{N}S(Au/YBCO)\approx 10^{-8}\Omega \cdot
cm^{2} $, $R_{N}S(Nb/Au)\approx 10^{-12}\Omega \cdot cm^{2}$ and $%
R_{N}S(Nb/YBCO)\approx 10^{-4}\Omega \cdot cm^{2}$. From the comparison with
the results presented in the table 1 for the junction Nb/Au/YBCO, we can
neglect the Nb/Au interface resistance, and take into account only Au/YBCO
interface resistance. $R_{d}(V)$ for Au/YBCO junctions show the similar
behavior as $R_{d}(V)$ for Nb/Au/YBCO.

Figure 4 presents the surface of bilayer structure Au/YBCO, measured with
the Atomic Force Microscope (AFM). One can see the strongly nonuniform
surface, which consists of Au granules with about 1$\mu m$ distance between
its. So, there is a good electrical contact in the parts of the film, where
Nb film covers Au granules on YBCO. But between the Au granules, where Nb
film has direct contact with the surface of YBCO, the oxygen depletion on
the surface of the structure causes the increase of the resistance. Direct
electrical contact Nb/YBCO has the specific resistance 3-4 orders of
magnitude higher than the Au/YBCO contact.

\section{Discussion}

We can consider the investigated structure as a parallel connection of
grains Nb/Au/YBCO and the parts of the structure with direct contact
Nb/YBCO. We suppose, that the current flows through Au/YBCO interface,
because of the big resistance Nb/YBCO interface and we can use the schematic
model, presented on the fig.5.: the superconducting electrode YBCO (S$_{d}$)
with the critical temperature of the superconducting transition $T_{cd}=87K$
and thickness 100-150nm; the damaged layer of YBCO (S$_{d}^{^{\prime }}$)
1-3nm thickness with suppressed superconducting order parameter; normal
metal layer 10-20nm thickness; the superconducting Nb counter electrode (S$%
^{^{\prime }}$) with $T_{c0}=9.2K$. The analogous model was suggested in [4]
for the evaluation of the electrophysical parameters in the system
Pb/Au/YBCO.

Let's evaluate the behavior of the superconducting order parameter in such
Nb/Au/YBCO structure. First, consider the Nb/Au interface. It's measured
resistance is low enough, so we can conclude, that superconducting Green
function (which is the characteristic of the amplitude of the interaction
potential of the superconducting carriers $\Phi $) and it's derivative are
continuous in the interface. Using calculations [12,13] we obtained the
value of the superconducting order parameter at the interface $\Delta
_{1}/e\approx $560$\mu V $, which is a bit smaller than in the bulk Nb. For
our evaluations we used the following values of electrophysical parameters
Nb and Au at $T=4.2K$: $\rho ^{Nb}l^{Nb}=4\cdot 10^{-12}\Omega \cdot cm^{2}$%
, $\xi ^{Nb}=7.3\cdot 10^{-7}cm$, $V_{F}^{Nb}=3\cdot 10^{7}cm/s$, $%
T_{c0}^{Nb}=9.2K$ and $\rho ^{Au}l^{Au}=8\cdot 10^{-12}\Omega \cdot cm^{2}$, 
$\xi ^{Au}=1\cdot 10^{-6}cm$, $V_{F}^{Au}=1.4\cdot 10^{8}cm/s$, where $%
V_{F}^{Nb,Au}$ - Fermi velocities and $l^{Nb,Au}$ - mean free paths of Nb
and Au correspondingly.

For the YBCO/Au interface we suppose, that it could be the 3nm thickness
damaged layer of YBCO $S_{d}^{^{\prime }}$ with the reduced critical
temperature (may be nonsuperconductive). Let's assume the small difference
in the coherence length $\xi _{c-YBCO}=\xi _{S_{d}^{^{\prime }}}=5\cdot
10^{-8}cm$ and that the specific resistance increases for the order of
magnitude [4] from $\rho _{c-YBCO}=1\cdot 10^{-4}\Omega \cdot cm$ to $\rho
_{S_{d}^{^{\prime }}}=1\cdot 10^{-3}\Omega \cdot cm$. We have the value of
the order parameter in YBCO near the Au/YBCO interface $\Delta
_{2}^{^{\prime }}/e\approx 140\mu V$ qualitatively. There is a potential
barrier with the low transparency $\bar{D}\sim 10^{-6}$ direct on the
Au/YBCO interface, which leads to the step of the order reduction in $\bar{D}
$ times:$\Delta _{2}=\Delta _{2}^{^{\prime }}\bar{D}$ [13]. We used the
theoretical evaluations for s-type pairing superconductors. But one can
conclude from the evaluations [10,14], that the behavior of the order
parameter in d-superconductor are qualitatively the same as for
s-superconductor if one of the main crystallographic directions of
d-superconductor is normal to the interface.

So, we can evaluate the amplitude value of the supercurrent through the
whole structure as in the junction superconductor-normal
metal-superconductor ($S_{d}^{^{\prime }}NS$), where values of the order
parameter on the interfaces are $\Delta _{2}/e\approx 0.14nV$ and $\Delta
_{1}/e\approx 560\mu V$. Than we use the theory developed for $SNS$%
-junctions. The thickness of the N-layer is of the same order of magnitude,
that it's coherence length, so we can neglect the decrease of the order
parameter in the N-layer and result $I_{c}R_{N}\approx \frac{\sqrt{\Delta
_{1}\Delta _{2}}}{e}=0.09\mu V$. Taking into account the value of the normal
resistance of the junction $R_{N}=10\Omega $, we get the value of $%
I_{c}\approx 0.009\mu A$, which is less than the fluctuation one for the
measurement system $I_{f}=1\mu A$ and could not be observed even in the case
of the dominant s-type component. If we have the pure $d_{x^{2}-y^{2}}$-
superconductor, the critical current is equal to zero from the symmetry of
the gap. The observed in [6-8] critical current in Pb/(Au,Ag)/YBCO
structures with higher values of $R_{N}S$ and junction areas, possibly,
caused by a soft etching in the Br-ethanol solution, which allows the
current flow through the contacts to a-b planes of YBCO. The transparencies
of the interfaces with normal metals in a-b planes of YBCO are 3 orders of
magnitude higher, than along c-axis ($R_{ab}S_{ab}<<R_{c}S_{c}$), but $%
S_{ab}<<S_{c}$. So, the normal resistance resulted from the parallel
connection of the interfaces along c-axis and a-b planes of YBCO is equal to 
$R_{N}S=R_{c}S$ if $S_{ab}/S_{c}\approx R_{ab}S_{ab}/(R_{c}S_{c}),
S=S_{ab}+S_{c}$. In the case of Nb, the oxygen depletion in a-b planes is
stronger than along c-axis, so the influence of a-b plane tunneling is
significantly reduced.

The important fact is the possible interaction both with YBCO and Au with
the formation of the superconducting alloy. In this case we have a
superconductor with the low enough critical temperature instead of the
normal metal, which could have a gap feature at low temperatures [6-8].

Let's discuss the figure3 - the $R_{d}(V)$ dependencies for the junctions at
4.2-100K temperature range. At $T<<T_{cd}$ the I-V curves correspond to the
SIN\ junction: the increase of $R_{d}$ below the gap value. But there are no
gap features of YBCO. These two facts correspond to a gapless
s-superconductor or d-superconductor with the nodes [14,16]. Calculations
[14] give the dependencies $R_{d}\sim \ln (T)$, $\ln (\mid \mid eV\mid
-\Delta \mid )$. Note, that for s-superconductor with gap we have $T^{-1/2}$%
, $((eV)^{2}-\Delta ^{2})^{-1/2}$.

For s-superconductors at $kT<<\Delta $, the number of excited quasiparticles
increases exponentially with the temperature, so $R_{d}(0)\sim \exp (-\Delta
/kT)$. It could be a great number of excited quasiparticles even at very low
temperatures $kT<<\Delta $ in d- superconductors due to the existence of
nodes with zero value of the order parameter. So, $R_{d}(0)$ must increase
slower with the decrease of temperature [14]. In our experiment we have the
linear $R(T)$ dependence with decrease of T (insert to a fig.2). The
dependence $R_{d}(V)$ corresponds qualitatively to the theoretical
calculations for d- superconductor [14].

One of the most interesting effects for d-superconductors is the existence
of two types of the bound states. The surface states at low energies caused
by a sign change of the order parameter for the reflected quasiparticles in
the a-b plane of d- superconductor [14-16]. It leads to a minimum value on $%
R_{d}(V)$ dependence at low $V$ I SIN junction, which was observed for a-b
plane current transport [6-8]. In our experiment we have no such effect due
to the quasiparticle momentum along c-axis of YBCO.

Another additional states could be present due to the suppression of the
order parameter of d-type superconductor at the interface for the directions
different from the main crystallographic ones or in the case of the
diffusive scattering [16]. These states are observed at nonzero energies and
weakly dependent from the temperature and the interface quality. One could
found these states as decrease of $R_{d}$ at the values of $V_{d}$ near the
gap of d-superconductor. The condition of the suppression of the order
parameter is realized in our experiment due to the degradation of YBCO
surface. Really, we found the weak minimum at $V\approx 15mV$ for all
measured samples, which didn't change it's position from the temperature.

\section{Conclusion}

Here we present the results of fabrication and experimental investigation of
\ the electron transport in Nb/Au/YBCO junctions along c-axis of YBCO. The
transparencies of the barriers are one order of magnitude higher the ones
from [6-8]. The evaluations on the basis of the proximity effect showed the
absence the critical current cased by the Au/YBCO interface potential
barrier. The $R_{d}(V)$dependencies are analogous to the case of SIN
junctions with gapless superconductor, in particular, the absence of the gap
feature for YBCO could correspond to d-superconductivity, due to the
existence of the nodes in the order parameter. $R(T)$ dependencies are also
in good agreement with the theory of d-superconductor. At $V\approx 15mV$
one could see the sharp decrease of the $R_{d}$, which could be caused by
the additional states at the interface of the d-type superconductor with the
normal metal.

The authors thank, Yu.S.Barash, D.A.Golubev, Z.G.Ivanov, M.Yu.Kupriyanov,
A.V.Zaitsev, for fruitful discussions; P.B.Mozhaev, D.Ertz, T.Henning for
the help during the experiment. This work was partially supported by the
program ''Modern problems in the physics of condensed state'', division
''Superconductivity'', Russian Foundation for Basic Research and the program
INTAS program of EU.

\section{References}

1.D.A.Wollman, D.J.Van Harlingen, W.C.Lee et.al., Phys.Rev.Lett. V.71, N.13,
pp.2134-2137 (1993).

2.C.C.Tsuei, J.R.Kirtley, C.C.Chi et.al., Phys.Rev.Lett.V.73, N.4,
pp.593-596 (1994).

3.H.Akoh, C.Camerlingo and S.Takada, Appl.Phys.Lett. V.56., N.15,
pp.1487-1489 (1990).

4.J.Yoshida, T.Hashimoto, S.Inoue et.al.Jpn.J.Appl.Phys.V.31, pp.1771-1777
(1992).

5.J.Lesueur, L.H.Greene, W.L.Feldmann et.al. Physica C V.191, pp.325-332
(1992).

6.A.G.Sun, A.Truscott, A.S.Katz et.al., Phys.Rev.B V.54, pp.6734-6741 (1996).

7.A.S.Katz, A.G.Sun, R.C.Dynes et.al., Appl.Phys.Lett. V.66, N1., pp.105-107
(1995).

8.J.Lesueur, M.Aprili, A.Goulon et.al., Phys.Rev.B V.55, N.6, pp.3398-3401
(1997).

9.J.R.Kirtley, K.A.Moler and D.J.Scarlapino, preprint cond-mat/9703067.

10.L.J. Buchholltz, M. Palumbo, D. Rainer, J.A. Sauls, J. of Low Temp. Phys.
V101, N5/6,pp.1079-1097,(1995).

11.A.V.Zaitsev, Sov.Phys.JETP V.86, N.5., pp.1742-1758 (1984).

12.M.Yu.Kupriyanov and K.K.Likharev, IEEE Trans. Magn.27, p.2400 (1991).

13.G.Deutscher, Physica C, V.185-189, pp.216-220 (1991).

14.Yu.S.Barash, A.V.Galaktionov and A.D.Zaikin, Phys.Rev.B V.52, N.1,
pp.665-682 (1995).

15.Y.Tanaka and S.Kashiwaya, Phys.Rev.B V.53, p.11957 (1996).

16.Yu.S.Barash, A.A.Svidzinsky and H.Bukhardt, Phys.Rev.B V.55, N.22,
pp.15282-15294 (1997).

17. P.V. Komissinski, G.A. Ovsyannikov, N.A. Tulina, V.V. Ryazanov. Current
transport for high-T superconductor/normal metal/ordinary superconductor. 31
National Russian Conference on Low Temperature Physics, Moscow, 1998 (in
Russian).

\section{Figure Captions}

Fig.1. The fabrication procedure of YBCO/Au/Nb junctions:a) Nb/Au/YBCO
trilayer structure deposition; b) definition of the area of the junction
with photolithography, ion and plasma chemical techniques; c) deposition of
insulating CeO$_{2}$ layer to prevent the electrical contacts in a-b plane
of YBCO; d)contact pads deposition; e)top view of the junction.

Fig2. The temperature dependence of the structure and of the 4$\mu m$ width
microbridge on the same substrate. The measurement current is 5$\mu A$. The
insert presents the $R(T)$ dependence on a larger scale at $T<T_{cd}$. The
resistance of microbridge is equal to zero at this temperature range.

Fig.3. Dependencies of the differential resistance of the structure at
different temperatures. Right axis shows the scale for T=91K.

Fig.4.Surface of the structure Au/YBCO, obtained by AFM.

Fig.5.Schematic dependence of the order parameter and amplitude of the pair
potential values in direction perpendicular to the junction plane.

\section{Table 1}

Parameters of fabricated junctions, measured at T=4.2K.

\begin{tabular}{cccccc}
Sample & $R_{d}(0),\Omega $ & $R_{N}(V),\Omega $ & $R_{N}S,10^{-6},\Omega
\cdot cm^{2}$ & $R_{d}(0)/R_{d}(V)$ & $\bar{D},10^{-6}$ \\ 
P9J2 & 12.2 & 7.0 & 4.5 & 1.7 & 4.8 \\ 
P9J3 & 9.8 & 6.0 & 3.8 & 1.6 & 5.6 \\ 
P10J2 & 10.5 & 5.9 & 3.8 & 1.8 & 5.6 \\ 
P10J3 & 10.6 & 5.9 & 3.8 & 1.8 & 5.6 \\ 
P11J2 & 4.9 & 4.2 & 2.7 & 1.2 & 7.9 \\ 
P11J3 & 5.2 & 3.6 & 2.3 & 1.4 & 9.3 \\ 
P12J2 & 2.4 & 2.0 & 1.3 & 1.2 & 16.7 \\ 
P13J2 & 7.2 & 3.5 & 2.2 & 2.1 & 9.5 \\ 
P13J3 & 7.5 & 6.6 & 4.2 & 1.1 & 5.1
\end{tabular}

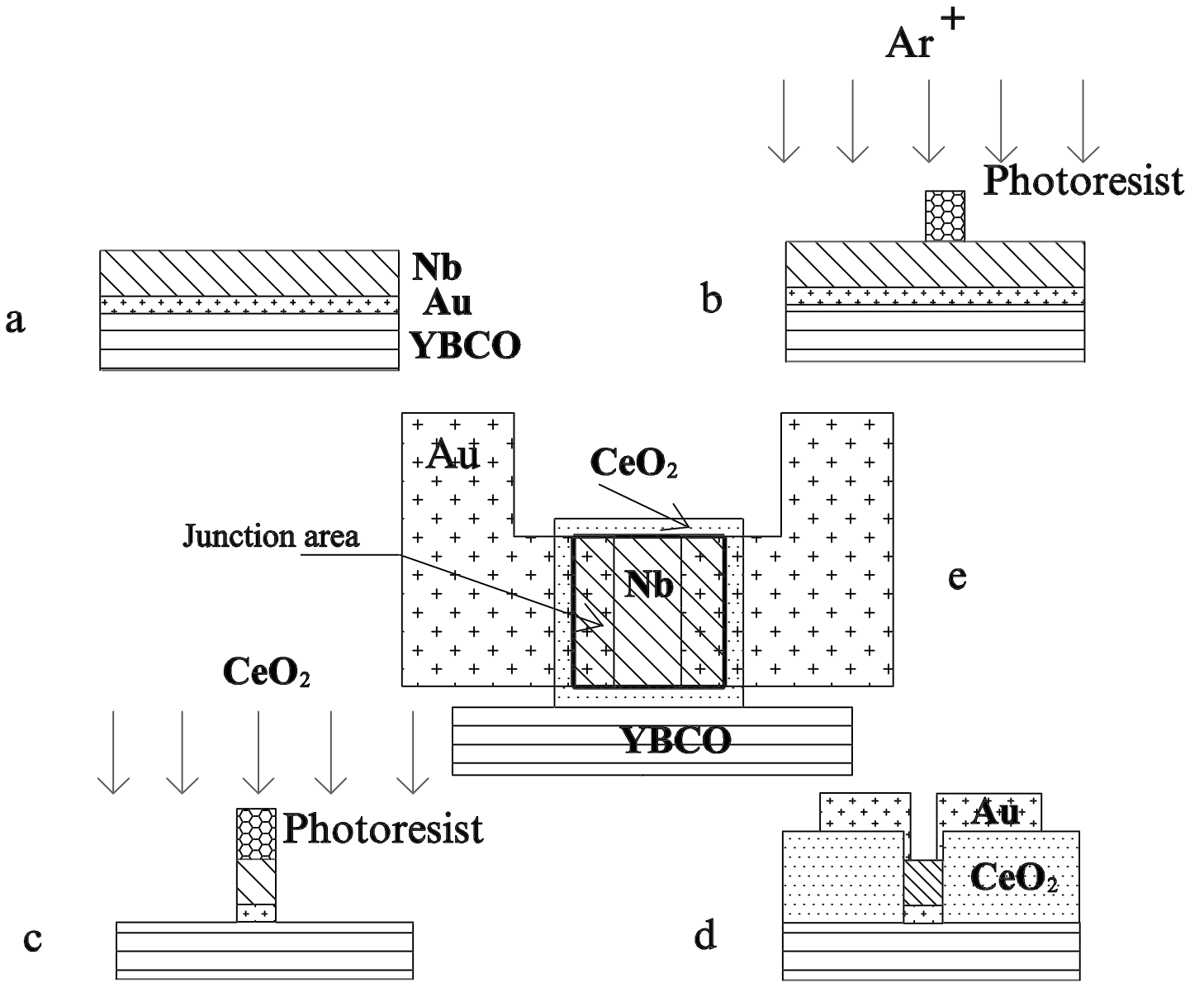
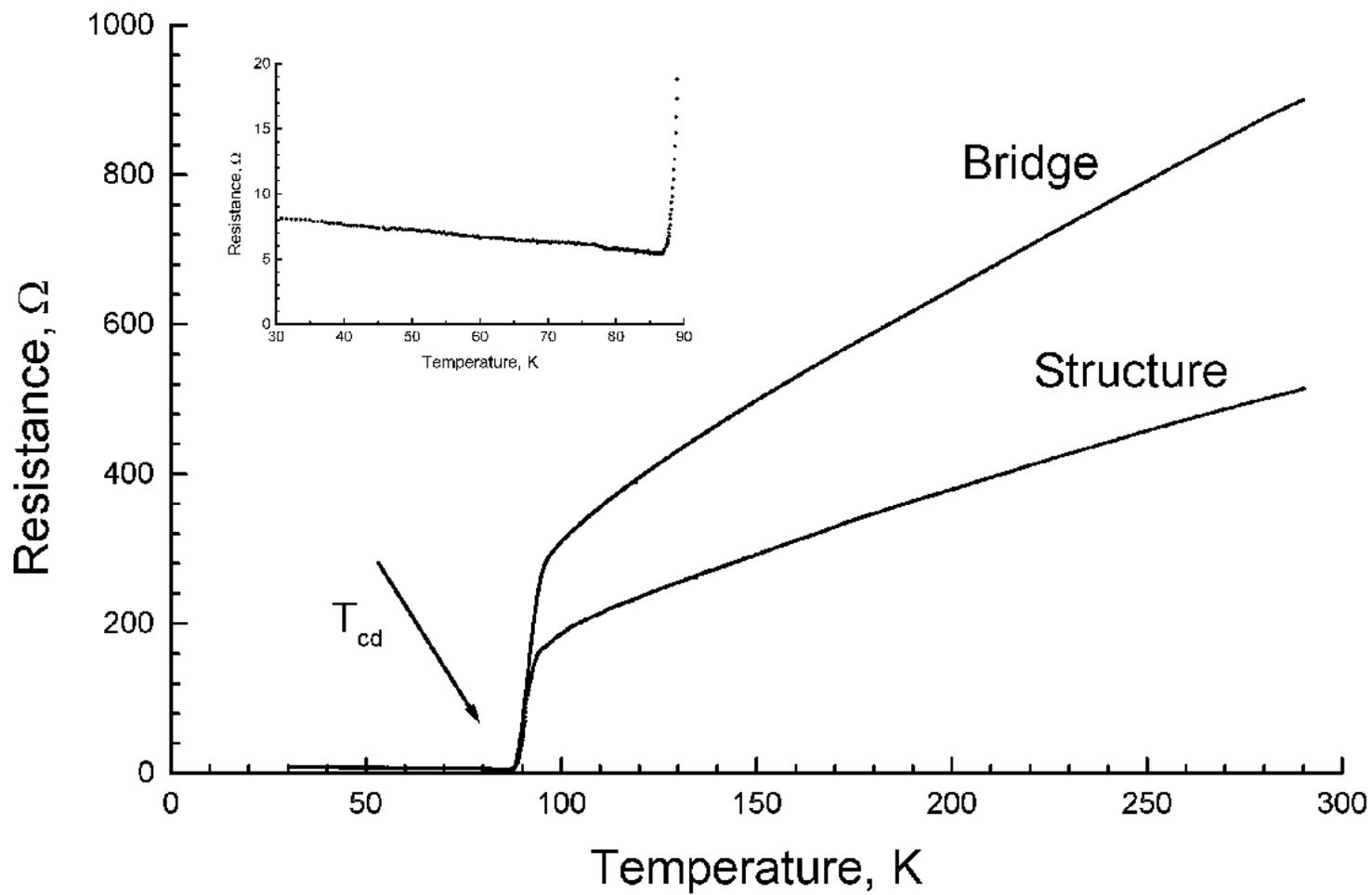
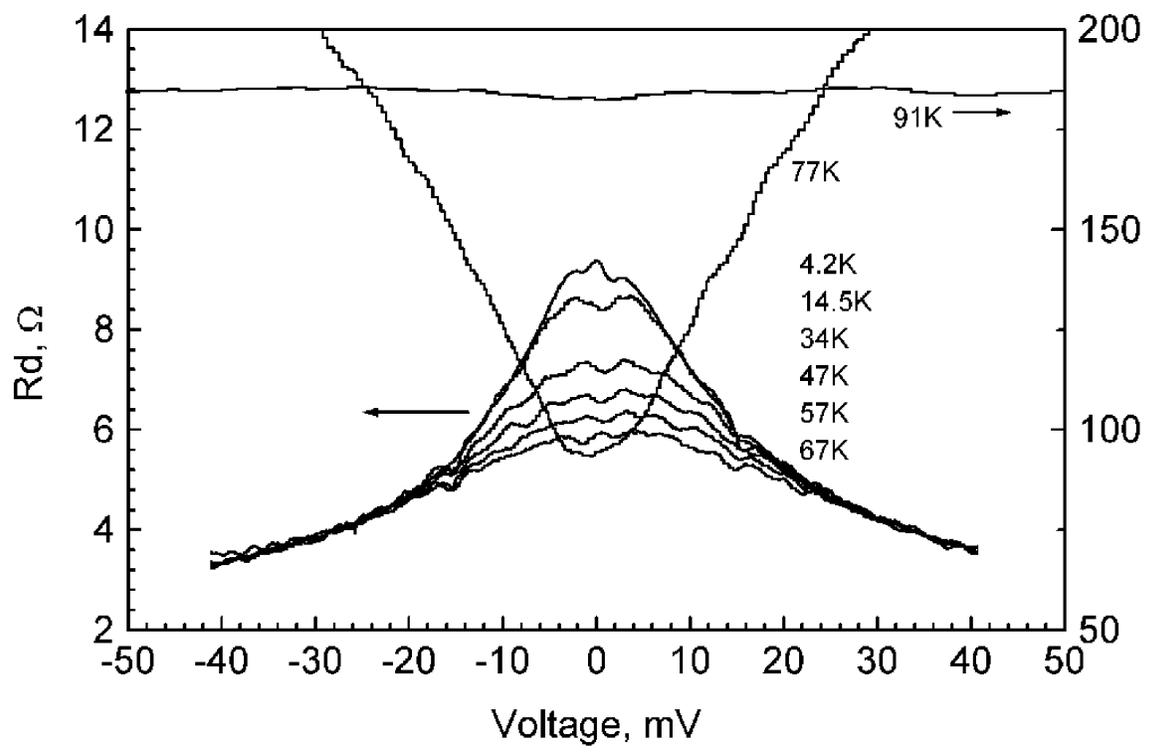
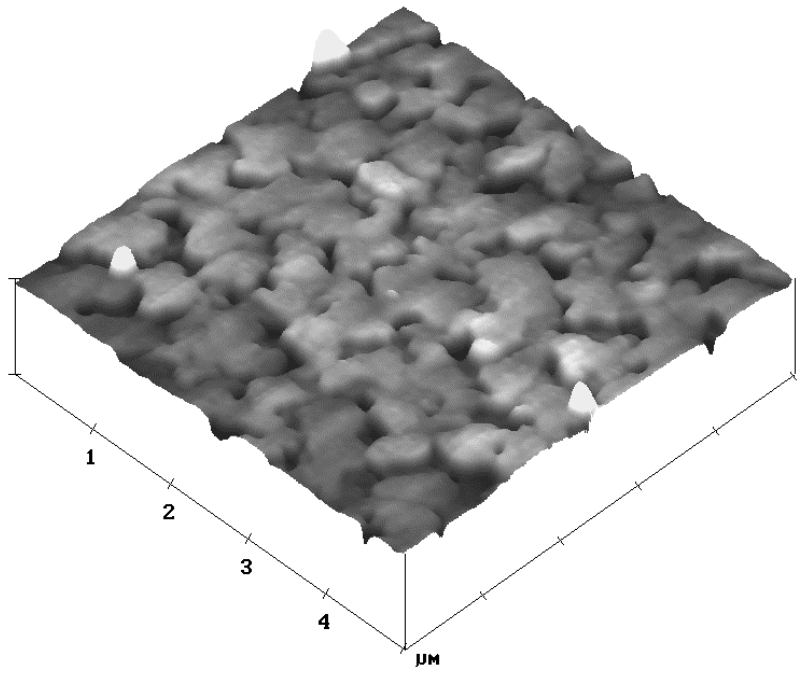
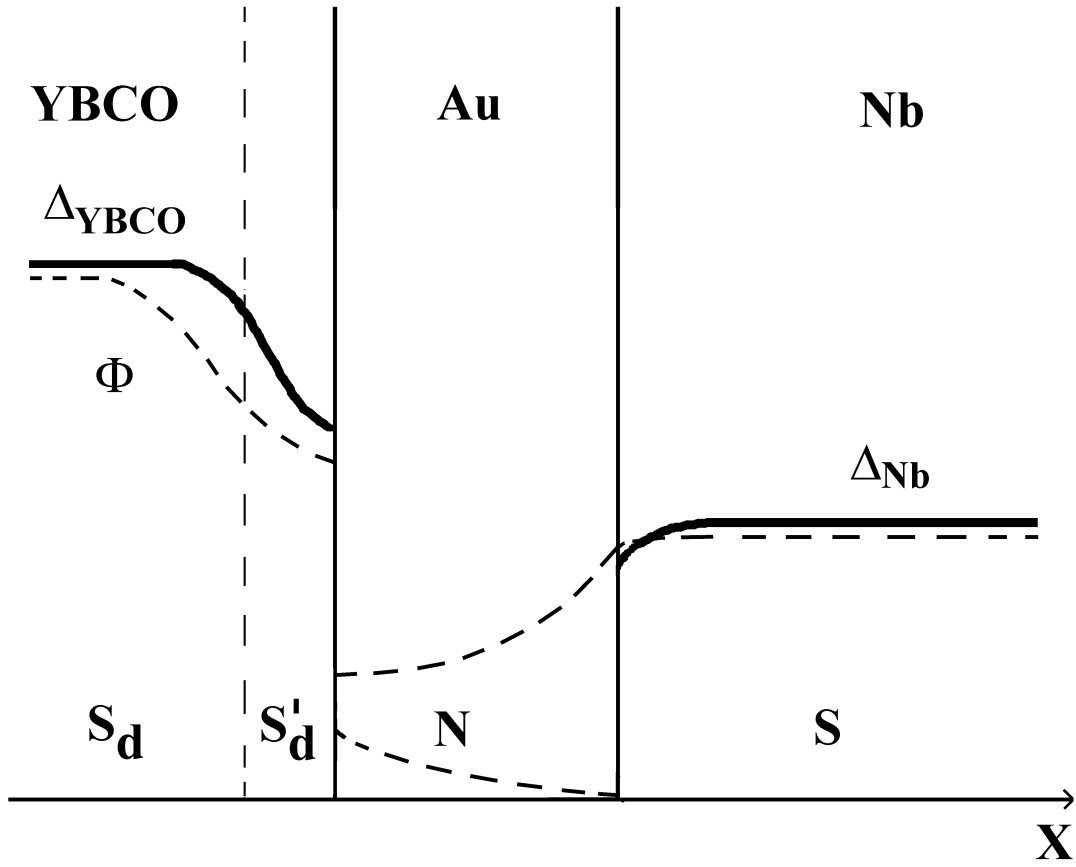
\end{document}